\documentclass[12pt]{article}
\usepackage{latexsym}
\usepackage{axodraw}
\usepackage{epsfig}
\usepackage{amsfonts}
\begin{document}

 \title{\bf Field Theory in Extra
Dimensions\thanks{Talk given at the conference "Supersymmetries
and Quantum Symmetries" (SQS'03) in memory of Prof.
V.I.Ogievetsky, July 2003, Dubna}}
\author{D.I.Kazakov}
\date{}
\maketitle \vspace{-0.8cm}
\begin{center}
{\it Bogoliubov Laboratory of Theoretical Physics, Joint
Institute for Nuclear Research, Dubna, Russia \\[0.1cm] and\\[0.1cm]
Institute for Theoretical and Experimental Physics, Moscow,
Russia}
\end{center}

\begin{abstract}
We analyze the possibility to construct a self-consistent   gauge
field theory in $D>4$. We first look for the cancellation of the
UV divergences in SUSY theories. Then, following the Wilson RG
approach, we study the RG equation for the gauge coupling in
perturbative and nonperturbative regimes. In the first case the
power low running is discussed. In the second case it is shown
that there exist the ultraviolet fixed point where the gauge
coupling is dimensionless in any space-time dimension. This fixed
point is nonperturbative and corresponds to scale invariant
theory. The same phenomenon also happens in supersymmetric theory
in D=6.
\end{abstract}

\section{Introduction}

The extra dimensional theories have already become popular for a
few years~\cite{ADD,RS}.  The main motivation for introducing
extra space dimensions comes from the string theory which prefers
to live in 26, 10 or 11 dimensions~\cite{GSW}. However, the string
theory is still far from being completed and the link to the lower
energy theory described by usual QFT is still missing. Staying in
the framework of QFT  one may wonder whether this extra
dimensional theory can be consistent in any sense. Since by
general power counting any interacting field theory (except for
$\phi^3$ in $D=6$) is nonrenormalizable, it looks hardly possible.

At the same time, following the concept of effective theory one
requires a consistency only up to a given order of perturbative
expansion, thus accepting the nonrenormalizable theories. Still,
the renormalizability or the possibility to consider, in
principle, a self-consistent theory served as a driving force in
construction of the Standard Model and should not be
underestimated. Whether or not this attitude may be applied to the
extra dimensional theories remains an open question.

In this talk I would like to concentrate on the following
questions:
\begin{itemize}
\item Can one construct a self-consistent QFT in $D > 4$ ?
\item Is it possible to get rid of UV divergences ?
\item Power law running: is it reliable ?
\item Nonperturbative fixed points: do they exist?
\end{itemize}

\section{UV divergences in SUSY theories for arbitrary $\mathbf D$.}

In principle, there is a chance that all the UV divergences cancel
each other, like it takes place in $N=4,\ 2$ and even $N=1$ SUSY
field theories in $D=4$~\cite{finite}, and one might have a
consistent theory. This possibility has been studied in the
literature~\cite{Ant}-\cite{K}.

Indeed, since one has an equivalence between extended
supersymmetry in D=4 and reduced supersymmetry in higher
dimensions, namely
$$N=4,\ D=4 \Leftrightarrow N=2,\ D=6 \Leftrightarrow N=1,\ D=10,$$
$$N=2, \ D=4 \Leftrightarrow N=1,\ D=6,$$
some cancellations take place. For example, total cancellation of
the quadratic and logarithmic UV divergences in $N=4$ $D=4$ SUSY
theory leads to the cancellation of the quartic and quadratic
divergences in $N=2$ $D=6$ theory and octic and sextic divergences
in $N=1$ $D=10$ case. Analogously  the condition which guarantees
the cancellation of the logarithmic divergences in $N=2$ $D=4$
theory works also for the quadratic divergences in $N=1$ $D=6$
case. The results are summarized  below (Here $C_A$ and $T_R$ are
the Casimir operators of the adjoint and arbitrary matter field
representations, respectively.)
\begin{table}[h]
  \centering
\begin{tabular}{|l|l|l|l|} \hline\hline
  $D$ & $N$ & UV Divergences  in one loop order & \\ \hline\hline
  $D=4$ & $N=1$ & $-11/3\ C_A+2/3\ C_A+2/3\ T_R+1/3\ T_R$& $=-(3C_A-T_R)$  \\ \hline
  \ $log\ \Lambda^2$ & $N=2$&$ -11/3\ C_A+2/3\ C_A+C_A+2\ T_R$ & $=-2(C_A-T_R)$  \\ \hline
   &   $N=4$&$-2\ C_A+2\ C_A$ &$=0$\\ \hline\hline
$D=6$ & $N=1$ & $-10/3\ C_A+4/3\ C_A+4/3\ T_R+2/3\ T_R$& $=-2(C_A-T_R)$  \\
\hline
 \ \ \  $ \Lambda^2$  & $N=2$& $-2\ C_A+2\ C_A$ & $=0$  \\ \hline\hline
$D=10$ &$N=1$&$-8/3\ C_A+8/3\ C_A$&$=0$\\ \hline
 \ \ \   $\Lambda^6$& & &\\ \hline\hline
\end{tabular}
 \caption{One loop UV divergences in SUSY gauge theories for arbitrary $D$ }\label{}
\end{table}

One can see that  the leading divergences indeed cancel each
other. However, this is not true anymore for the logarithmic
divergences. Strictly speaking they are not gauge invariant and to
get the gauge invariant statement one has to go on shell. Then at
lower orders the divergences indeed cancel~\cite{T,MS,K}, but in
higher orders they may well appear being unprotected by any
symmetry~\cite{HS}. Indeed, it has been checked by explicit
calculation in components~\cite{T,MS} that $D=6\ N=1$ SUSY gauge
theory is on-shell finite up to two loops. However, within the
(constrained) superfield formalism it is possible to show that the
nonvanishing invariants  in higher loops exist~\cite{HS}.

Thus, the theory remains {\it perturbatively nonrenormalizable}.

\section{Wilsonian RG in $\mathbf{D>4}$.}

We  now  look for the alternative possibilities to construct a
viable higher dimensional theory. We follow the so-called Wilson
Renormalization Group approach~\cite{Wilson}.

Consider first the usual gauge theory in $D$ dimensions
\begin{equation}\label{l}
  {\cal L}=-\frac 14 Tr F^2_{\mu\nu}, \ \ \
  F_{\mu\nu}=\partial_\mu A_\nu - \partial_\nu A_\mu + g
  [A_\mu,A_\nu].
\end{equation}
The fields and the coupling have the following canonical
dimensions:
$$[A]=\frac{D-2}{2}\ ,\ \  \ \ [F]=\frac D2 \ ,\ \  \ \ [g]=2-\frac D2. $$
Thus, $D=4$ is the critical dimension: the coupling  is
dimensionless, the operators are marginal and the theory is
renormalizable in a usual sense.

To go beyond the critical dimension we follow the standard
approach~\cite{H} (see also Ref.\cite{ZJ}) based on dimensional
regularization and analytical continuation. Consider the
dimensionless quantity
$$\tilde g \equiv g \mu^{D/2-2} \ \ \ \ \Rightarrow \ \ \ \ [\tilde g] = 0,$$
where $\mu$ is some scale, and expresses the bare coupling in
terms of a renormalized one
\begin{equation}\label{b}
  g_B=\mu^{2-D/2}\tilde{g}Z_g(\tilde{g}),
\end{equation}
where the renormalization constant $Z_g$ depends on $D$ and in the
minimal subtraction scheme contains only the pole terms in
$(D-D_c)$ with $D_c=4$ in this case.

The crucial point here is that {\it the renormalization constant
$Z_g$ depends on $\tilde{g}$ and does not contain the infinite
number of higher dimensional operators that may appear at
$D>D_c$}. In Wilson approach these operators are irrelevant while
going toward the {\it infrared} direction and may be
ignored~\cite{Wilson}.

Differentiating then eq.(\ref{b}) with respect to a scale keeping
$g_B$ fixed one gets
\begin{equation}
  0=(2-\frac D2)\tilde{g}+\beta(\tilde g)-\tilde g
  \gamma_g(\tilde{g}),
\end{equation}
where as usual $\beta(\tilde g)=\mu \frac{d}{d\mu} \tilde
g|_{g_B}$ and $\gamma_g=-\mu \frac{d}{d\mu} ln Z_g |_{g_B}$. This
leads to the following RG equation for the coupling
\begin{equation}\label{rg}
 \mu\frac{d}{d\mu}\tilde g =   \beta(\tilde g)=\tilde g(\frac D2-2+\gamma_g).
\end{equation}
In general $\gamma_g(g)$, and hence $\beta(g)$,  may depend on $D$
being finite while $D\to D_c$. However, in the MS-scheme this
dependence is absent and $\gamma_g$ can be calculated directly in
the critical dimension. We use this advantage since while the
$\beta$ function and the anomalous dimensions are scheme
dependent, the value of the anomalous dimension at the critical
point is universal. This means that it can be calculated in any
scheme. It is useful to proceed in the background field gauge
where $\gamma_g =\frac 12 \gamma_A$ the latter being the gauge
field anomalous dimension.

\section{A perturbative solution}

Consider now the perturbative solution to eq.(\ref{rg}). For this
purpose we take the one loop expression for
$\gamma_A=b\tilde{g}^2$. Then the solution
 looks like
\begin{equation}\label{pt}
  \frac{1}{\tilde{g}^2}=\frac{1}{\tilde{g}_0^2}\left(\frac{\mu^2}{Q^2}\right)^\varepsilon
  +\frac{b}{2\varepsilon}\left[\left(\frac{\mu^2}{Q^2}\right)^\varepsilon-1\right],
  \ \ \ \ \varepsilon = \frac{D}{2}-2,
\end{equation}
and exhibits the power law running. When $D\to 4$ or
$\varepsilon\to 0$ one comes back to the usual logarithmic
behaviour
\begin{equation}\label{log}
\frac{1}{\tilde{g}^2}=\frac{1}{\tilde{g}_0^2}+\frac
b2\log\frac{\mu^2}{Q^2}.
\end{equation}

Eq.(\ref{pt}), being obtained though on the basis of Kaluza-Klein
approach, has been considered as the way to the low scale
unification in Grand Unified Theories~\cite{DDG}. Indeed, if one
assumes that at some energy scale the extra dimension comes into
play, one has to switch at this scale from the log running in
$D=4$ into the power running in $D>4$ and gets the lower scale
unification as shown in Fig.{\ref{un}.
\begin{figure}[ht]
  \centering
  \leavevmode
\epsfxsize=12cm \epsfysize=8cm
 \epsffile{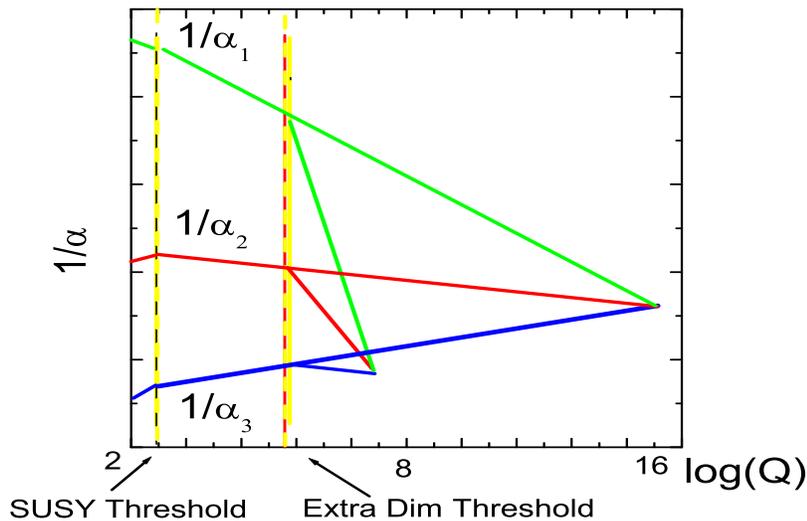}\vspace{-0.5cm}
  \caption{Power law versus logarithmic unification of the couplings}\label{un}
\end{figure}

\noindent Note that the very fact of unification of the three
curves according to eq.(\ref{pt}) does not depend on $\varepsilon$
and if the curves unify for $D=4$ they will do it for any $D$.

This very appealing picture, however, suffers one substantial
problem. Strictly speaking this power law running is valid if
going the {\it infrared} direction. Just in this case one can
ignore the infinite number of higher dimensional operators which
are irrelevant here. When going ultraviolet direction they all
become relevant and eq.(\ref{rg}) with a single coupling is no
more valid. This irreversibility of the RG equations in the Wilson
approach is essential and can not be ignored. One may hope to
overcome this difficulty in some underlying theory, like the
string one, however, the result is unclear. Therefore, to my mind,
the advocated power law running of the couplings with the low
scale unification in the UV region {\it can not} be considered as
reliable so far.

\section{Nonperturbative fixed point}

Consider now the nonperturbative solution to eq.(\ref{rg}). It
has two fixed points
$$\begin{array}{l}
1) \ \ \tilde g=0 \ \ \rightarrow g=0, \ \ \gamma_A=0, \\
2) \ \ \tilde g=g^*, \ \ \gamma_A=4-D.
\end{array}$$
The first one is trivial, this is the so-called Gaussian fixed
point, it is perturbative. The second one is nonperturbative, it
is the so-called Wilson-Fisher fixed point~\cite{Wilson}. The
anomalous dimension here is not small, it is integer. It is
achieved at the value of the coupling which is unknown, though the
value of the anomalous dimension is known {\it exactly}. Since the
anomalous dimension in gauge theories, contrary to the scalar
case, is negative, the fixed point of the second kind exists for
$D>4$. (see Fig.\ref{fp}).
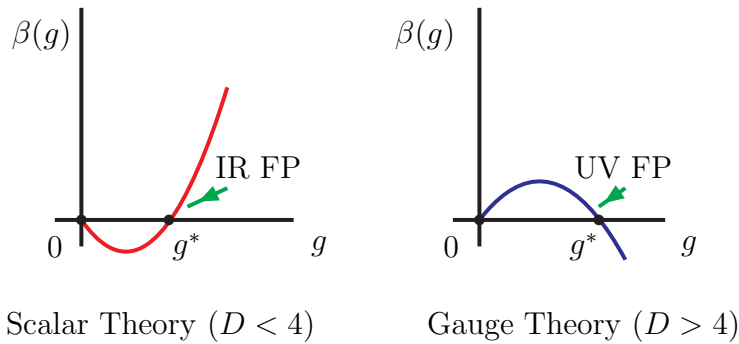
\begin{figure}[ht]
\begin{center}\SetWidth{1.5}
\begin{picture}(200,130)(30,-30)\Text(-5,80)[]{$\beta(g)$}\Text(100,0)[]{$g$}
\Text(0,0)[]{$0$}\Text(50,0)[]{$g^*$}
\Line(0,10)(90,10)\Line(10,0)(10,90)
\SetColor{Red}\Curve{(10,10)(20,-0)(65,60)}\SetColor{Black}
\Vertex(10,10){2}\Vertex(43,10){2}
\Line(150,10)(230,10)\Line(160,0)(160,90)\Text(140,80)[]{$\beta(g)$}\Text(240,0)[]{$g$}
\SetColor{Blue}\Curve{(160,10)(170,20)(215,-5)}\SetColor{Black}
\Text(150,0)[]{$0$}\Text(200,0)[]{$g^*$}\Vertex(160,10){2}\Vertex(205,10){2}
\Text(40,-30)[]{Scalar Theory\ ($D<4$)}\Text(200,-30)[]{Gauge
Theory\ ($D>4$)} \Text(77,30)[]{IR FP}\Text(215,30)[]{UV FP}
\SetColor{Green}
\ArrowLine(65,22)(50,15)\ArrowLine(215,22)(205,15)
\end{picture}\caption{The fixed points in scalar and gauge theories}\label{fp}
\end{center}
\end{figure}

Such a fixed point in a gauge theory within the
$\epsilon$-expansion has been  advocated in ref.\cite{P}. Some
additional supporting arguments in favour of the fixed point in 5
dimensions come from the lattice calculations~\cite{Lat}. At last,
there is also very useful analogy between the gauge theory and the
two dimensional nonlinear sigma-model. The latter has  a fixed
point in the leading order~\cite{PS} which is also true within the
1/N expansion performed directly in three dimensions~\cite{A}.

 Consider the properties of the fixed point $\# 2$.
Let us calculate the dimensions. One has for the field
$$[A]=\frac{D-2}{2}+\frac
12\gamma_A=\frac{D-2}{2}+\frac{4-D}{2}=1$$ in any $D$. To
calculate the dimension of the coupling, one has to consider the
vertex $g\partial A[A,A]$ which gives
$$D=[g]+1+3[A]+\gamma_V.$$\
Since $\gamma_V=-\gamma_A$ in the background gauge, one obtains
$$[g]=D-4-\gamma_V=D-4+\gamma_A=0  \ \ \ \ \mbox{in any D}\ \ \mathbf{!}$$

Thus, one has a dimensionless coupling at the fixed point that
means renormalizability. {\it The theory at the fixed point is
perturbatively nonrenormalizable, but nonperturbatively
renormalizable!}~\cite{S,S2}). The existence of a renormalizable
field theory beyond PT relies, in the sense of statistical
physics, on the existence of a fixed point (see e.g
Ref.~\cite{ZJ}, p.549).

Since the full  dimension of the field is known, it is possible to
calculate the behaviour of the propagator at the fixed point. One
has
\begin{equation}\label{prop}
\widehat{AA} \sim \frac{\displaystyle 1}{\displaystyle (x^2)^1}
\Rightarrow \int\frac{\displaystyle d^Dx e^{ipx}}{\displaystyle
(x^2)}\sim \frac{\displaystyle 1}{\displaystyle
(p^2)^\frac{D-2}{2}}
\end{equation}
Thus, for instance, for $D=6$ at the non-Gaussian fixed point the
propagator behaves like $1/p^4$, i.e. much faster than in the
usual case.

One can try to construct an effective Lagrangian that takes into
account the anomalous dimensions calculated above. In $D=6$, as it
is suggested by the one-loop calculation and the behaviour of the
propagator, it is
\begin{equation}\label{eff}
  {\cal L}_{eff}\sim \ Tr (D_\mu F_{\mu\nu})^2.
\end{equation}

The effective Lagrangian (\ref{eff}) has a remarkable property: it
is {\it scale} invariant. Earlier I assumed that it might also  be
{\it conformal} invariant~\cite{K2}, though conformal invariance
does not necessarily follows from the scale one~\cite{Vas} and has
to be checked. If it is true, then this will essentially constrain
the Green functions allowing the non-perturbative information of
their properties~\cite{Tod}.

\section{Nonperturbative fixed point in  SUSY theories}

A similar phenomenon takes place in SUSY gauge theories. Here,
however, we are faced with the problem: supersymmetry does not
exist in any dimension. Indeed, from the requirement of equal
number of the fermionic and bosonic degrees of freedom for the
gauge field and its superpartner one gets
\begin{equation}\label{df}
  D-2=2^{[D/2]-1(2)}
\end{equation}
with the solution $D=4,6,10$ and possible modification for the odd
values of $D$. Moreover, as was already mentioned, higher
dimensional supersymmetry is equivalent to extended one in lower
dimensions.

Therefore, when considering supersymmetric theory in one of the
possible higher dimensions and going to the critical dimension
following the minimal subtraction procedure,  one has to keep
track of degrees of freedom. For instance, choosing N=1 D=6 theory
and going to D=4 one has to take N=2 SUSY in D=4, choosing N=1
D=10 theory one has to take N=4 D=4 SUSY, etc.

The number of possibilities, hence, is very limited. One has an
extended SUSY theory in D=4 with the single coupling: all the
Yukawa couplings are equal to the gauge one due to extended
supersymmetry.

Remind that in 4 dimensions N=2 SUSY theory (in supefield
formalism) has only one loop UV divergences, while N=4 SUSY theory
is totally finite (see Table 1). This means that eq.(\ref{rg}) in
SUSY case is essentially simplified. One has
\begin{equation}\label{1l}
  \gamma_A=b\tilde{g}^2\ \ \mbox{for}\ \  D=6\ \ \  \mbox{and} \ \ \gamma_A=0
  \ \ \mbox{for} \ \ D=10.
\end{equation}
Consequently, the nonperturbative fixed  point defined by the
condition $\gamma_A=4-D$  exists in $D=6$ when
$$\tilde{g}^2={g^*}^2=\ -\frac{2}{b}, \ \ \ b<0,$$
and does not exist in $D=10$ due to vanishing of $\gamma_A$ in
this case. Taking the value of $b$ from the Table 1 one finds
$$b \sim -2(C_A-T_R)<0 \ \ \ \Rightarrow \ \ T_R < C_A.$$
Thus, if the number of hypermultiplets is not big, the fixed point
exists. In particular it exists in pure gauge case when $T_R$=0.
Remarkably, that in SUSY case one knows not only the anomalous
dimension, but the critical coupling as well.

 At this fixed point a theory possess all the
properties mentioned above. It is perturbatively
nonrenormalizable, but nonperturbati\-ve\-ly renormalizable and
scale invariant. The effective action should be the SUSY
generalization of that of eq.(\ref{eff}).

There is another subtlety with supersymmetry. While supersymmetric
gauge theory exists for $D\leq 10$, it is known that
superconformal algebra is only possible for $D\leq 6$~\cite{Nahm}.
Earlier~\cite{K2}  I claimed one can get the fixed point in SUSY
gauge theory in any D. However, the present analysis shows the
existence of the nontrivial fixed point in $D=6$, and may be in
D=5, but not in $D=10$, that matches  the above
statement~\cite{Nahm} and resolves the contradiction.

These observed fixed points may be related to those originated
from the  string dynamics  for D=5 and 6 ~\cite{SW}. We use here
the more familiar language that is close to statistical physics
and critical phenomena. In a sense we give an explicit example of
a local field theory  with non-trivial fixed point thus
strengthening the claim (based on string theory) that exist field
theories that flow to non-trivial fixed points in more than 3
dimensions~\cite{SW}.

\section{Conclusion}

Summarizing the analysis of the gauge and SUSY field theories in
higher dimensions from the point of view of their
renormalizability and consistency, we come to the following
conclusions
\begin{enumerate}
  \item[-] Perturbative finiteness in $D>4$ seems not to be valid;
  \item[-] Within the Wilson RG approach one can write the equation
   for the couplings in $D>4$ which exhibits the power law running in the
    infrared direction, but not in the ultraviolet;
  \item[-] In this approach there exist the nontrivial
  nonperturbative fixed point which may lead to nonperturbative
  renormalizability;
  \item[-] At the fixed point the theory possesses the scale
  invariance, and the anomalous dimensions are known exactly;
\item[-] The same phenomenon happens in SUSY theories in D=6
(D=5);
   \item[-]  It is quite possible that at the fixed point N=1 D=6
    SUSY gauge theory, like the N=4 D=4 one, is also conformal invariant.
\end{enumerate}

\section*{Acknowledgements}
 Financial support from RFBR grant \#
02-02-16889 and the grant of Russian Ministry of Industry, Science
and Technologies \# 2339.2003.2 is kindly acknowledged.

\end{document}